

\def\singlespace{\normalbaselines}
\def\oneandahalfspace{\baselineskip=1.15\normalbaselineskip plus 1pt
\lineskip=2pt\lineskiplimit=1pt}

\def\np{\vfill\eject}
\def\nl{\hfil\break}

\def\nofirstpagenoten{\nopagenumbers\footline={\ifnum\pageno>1\tenrm
\hss\folio\hss\fi}}
\def\nofirstpagenotwelve{\nopagenumbers\footline={\ifnum\pageno>1\twelverm
\hss\folio\hss\fi}}
\def\leaderfill{\leaders\hbox to 1em{\hss.\hss}\hfill}
\def\ft#1#2{{\textstyle{{#1}\over{#2}}}}
\def\frac#1/#2{\leavevmode\kern.1em
\raise.5ex\hbox{\the\scriptfont0 #1}\kern-.1em/\kern-.15em
\lower.25ex\hbox{\the\scriptfont0 #2}}
\def\sfrac#1/#2{\leavevmode\kern.1em
\raise.5ex\hbox{\the\scriptscriptfont0 #1}\kern-.1em/\kern-.15em
\lower.25ex\hbox{\the\scriptscriptfont0 #2}}


\parindent=20pt
\def\narrow{\advance\leftskip by 40pt \advance\rightskip by 40pt}

\def\AB{\bigskip
        \centerline{\bf ABSTRACT}\medskip\narrow}
\def\nonarrower{\advance\leftskip by -40pt\advance\rightskip by -40pt}
\def\AE{\bigskip\nonarrower}

\def\boxit#1{\vbox{\hrule\hbox{\vrule\kern3pt
        \vbox{\kern3pt#1\kern3pt}\kern3pt\vrule}\hrule}}

\def\gtorder{\mathrel{\raise.3ex\hbox{$>$}\mkern-14mu
             \lower0.6ex\hbox{$\sim$}}}
\def\ltorder{\mathrel{\raise.3ex\hbox{$<$}|mkern-14mu
             \lower0.6ex\hbox{\sim$}}}
\def\dalemb#1#2{{\vbox{\hrule height .#2pt
        \hbox{\vrule width.#2pt height#1pt \kern#1pt
                \vrule width.#2pt}
        \hrule height.#2pt}}}

\font\fourteentt=cmtt10 scaled \magstep2
\font\fourteenbf=cmbx12 scaled \magstep1
\font\fourteenrm=cmr12 scaled \magstep1
\font\fourteeni=cmmi12 scaled \magstep1
\font\fourteenss=cmss12 scaled \magstep1
\font\fourteensy=cmsy10 scaled \magstep2
\font\fourteensl=cmsl12 scaled \magstep1
\font\fourteenex=cmex10 scaled \magstep2
\font\fourteenit=cmti12 scaled \magstep1
\font\twelvett=cmtt10 scaled \magstep1 \font\twelvebf=cmbx12
\font\twelverm=cmr12 \font\twelvei=cmmi12
\font\twelvess=cmss12 \font\twelvesy=cmsy10 scaled \magstep1
\font\twelvesl=cmsl12 \font\twelveex=cmex10 scaled \magstep1
\font\twelveit=cmti12
\font\tenss=cmss10
 
 \font\ninebf=cmbx7 scaled \magstep1
\font\ninerm=cmr7 scaled \magstep1 \font\ninei=cmmi7 scaled \magstep1
\font\ninesy=cmsy7 scaled \magstep1 
\font\eightrm=cmr7 scaled 1140 
 
\font\sevenbf=cmbx7 \font\sevenrm=cmr7 \font\seveni=cmmi7
\font\sevensy=cmsy7 

\catcode`@=11
\newskip\ttglue
\newfam\ssfam

\def\fourteenpoint{\def\rm{\fam0\fourteenrm}
\textfont0=\fourteenrm \scriptfont0=\tenrm \scriptscriptfont0=\sevenrm
\textfont1=\fourteeni \scriptfont1=\teni \scriptscriptfont1=\seveni
\textfont2=\fourteensy \scriptfont2=\tensy \scriptscriptfont2=\sevensy
\textfont3=\fourteenex \scriptfont3=\fourteenex \scriptscriptfont3=\fourteenex
\def\it{\fam\itfam\fourteenit} \textfont\itfam=\fourteenit
\def\sl{\fam\slfam\fourteensl} \textfont\slfam=\fourteensl
\def\bf{\fam\bffam\fourteenbf} \textfont\bffam=\fourteenbf
\scriptfont\bffam=\tenbf \scriptscriptfont\bffam=\sevenbf
\def\tt{\fam\ttfam\fourteentt} \textfont\ttfam=\fourteentt
\def\ss{\fam\ssfam\fourteenss} \textfont\ssfam=\fourteenss
\tt \ttglue=.5em plus .25em minus .15em
\normalbaselineskip=16pt
\abovedisplayskip=16pt plus 4pt minus 12pt
\belowdisplayskip=16pt plus 4pt minus 12pt
\abovedisplayshortskip=0pt plus 4pt
\belowdisplayshortskip=9pt plus 4pt minus 6pt
\parskip=5pt plus 1.5pt
\setbox\strutbox=\hbox{\vrule height12pt depth5pt width0pt}
\let\sc=\tenrm
\let\big=\fourteenbig \normalbaselines\rm}
\def\fourteenbig#1{{\hbox{$\left#1\vbox to12pt{}\right.\n@space$}}}

\def\twelvepoint{\def\rm{\fam0\twelverm}
\textfont0=\twelverm \scriptfont0=\ninerm \scriptscriptfont0=\sevenrm
\textfont1=\twelvei \scriptfont1=\ninei \scriptscriptfont1=\seveni
\textfont2=\twelvesy \scriptfont2=\ninesy \scriptscriptfont2=\sevensy
\textfont3=\twelveex \scriptfont3=\twelveex \scriptscriptfont3=\twelveex
\def\it{\fam\itfam\twelveit} \textfont\itfam=\twelveit
\def\sl{\fam\slfam\twelvesl} \textfont\slfam=\twelvesl
\def\bf{\fam\bffam\twelvebf} \textfont\bffam=\twelvebf
\scriptfont\bffam=\ninebf \scriptscriptfont\bffam=\sevenbf
\def\tt{\fam\ttfam\twelvett} \textfont\ttfam=\twelvett
\def\ss{\fam\ssfam\twelvess} \textfont\ssfam=\twelvess
\tt \ttglue=.5em plus .25em minus .15em
\normalbaselineskip=14pt
\abovedisplayskip=14pt plus 3pt minus 10pt
\belowdisplayskip=14pt plus 3pt minus 10pt
\abovedisplayshortskip=0pt plus 3pt
\belowdisplayshortskip=8pt plus 3pt minus 5pt
\parskip=3pt plus 1.5pt
\setbox\strutbox=\hbox{\vrule height10pt depth4pt width0pt}
\let\sc=\ninerm
\let\big=\twelvebig \normalbaselines\rm}
\def\twelvebig#1{{\hbox{$\left#1\vbox to10pt{}\right.\n@space$}}}

\def\tenpoint{\def\rm{\fam0\tenrm}
\textfont0=\tenrm \scriptfont0=\sevenrm \scriptscriptfont0=\fiverm
\textfont1=\teni \scriptfont1=\seveni \scriptscriptfont1=\fivei
\textfont2=\tensy \scriptfont2=\sevensy \scriptscriptfont2=\fivesy
\textfont3=\tenex \scriptfont3=\tenex \scriptscriptfont3=\tenex
\def\it{\fam\itfam\tenit} \textfont\itfam=\tenit
\def\sl{\fam\slfam\tensl} \textfont\slfam=\tensl
\def\bf{\fam\bffam\tenbf} \textfont\bffam=\tenbf
\scriptfont\bffam=\sevenbf \scriptscriptfont\bffam=\fivebf
\def\tt{\fam\ttfam\tentt} \textfont\ttfam=\tentt
\def\ss{\fam\ssfam\tenss} \textfont\ssfam=\tenss
\tt \ttglue=.5em plus .25em minus .15em
\normalbaselineskip=12pt
\abovedisplayskip=12pt plus 3pt minus 9pt
\belowdisplayskip=12pt plus 3pt minus 9pt
\abovedisplayshortskip=0pt plus 3pt
\belowdisplayshortskip=7pt plus 3pt minus 4pt
\parskip=0.0pt plus 1.0pt
\setbox\strutbox=\hbox{\vrule height8.5pt depth3.5pt width0pt}
\let\sc=\eightrm
\let\big=\tenbig \normalbaselines\rm}
\def\tenbig#1{{\hbox{$\left#1\vbox to8.5pt{}\right.\n@space$}}}
\let\rawfootnote=\footnote \def\footnote#1#2{{\rm\parskip=0pt\rawfootnote{#1}
{#2\hfill\vrule height 0pt depth 6pt width 0pt}}}

\overfullrule=0pt
\twelvepoint
\def\sbullet{\raise.2em\hbox{$\scriptscriptstyle\bullet$}}
\nofirstpagenotwelve
\hsize=16.5 truecm
\baselineskip 15pt

\def\ft#1#2{{\textstyle{{#1}\over{#2}}}}
\def\sss{\scriptscriptstyle}
\def\a{\alpha_0}

\def\del{\partial}

\oneandahalfspace
\rightline{CTP TAMU-57/92}
\rightline{EFI-92-39}
\rightline{July, 1992}

\vskip 2truecm
\centerline{\bf A Note on $N=2$ Superstrings}
\vskip 1.5truecm
\centerline{J. R. Bie\'nkowska}
\medskip
\centerline{\it Enrico Fermi Institute}
\centerline{\it The University of Chicago}
\centerline{\it 5640 South Ellis Ave.}
\centerline{\it Chicago, IL 60637-1433, U.S.A.}
\bigskip
\centerline{and}
\bigskip
\centerline{H. Lu}
\medskip
\centerline{\it Center for Theoretical Physics}
\centerline{\it Texas A \& M University}
\centerline{\it College Station, TX 77843, U.S.A.}
\bigskip

\AB\singlespace


      In this note we investigate the generalised critical $N=2$ superstrings
in $(1,2p)$ spacetime signature.  We calculate the four-point functions for
the tachyon operators of these theories.  In contrast to the usual $N=2$
superstring in $(2,2)$ spacetime, the four-point functions do not vanish. The
exchanged particles of the four-point function are  included in the physical
spectrum of the corresponding theory and have vanishing fermion charge.

\vskip 2truecm
\centerline{\tenpoint Available from hep-th/9207062}

\np
\AE\oneandahalfspace
\bigskip\noindent
{\bf 1. Introduction}
\bigskip\bigskip

       $N=2$ superstrings have received a considerable amount of attention
recently [1,2,3].  They are described by two-dimensional conformal field
theories with local $N=2$ super-Virasoro symmetry algebra.  They have the
maximal number of local supersymmetries which allow string theories to have
a positive critical central charge.  The critical value of  central charge
for the $N=2$ super-Virasoro algebra is $c=6$.  Thus it can be realised by
two $N=2$ superfields, which then leads to the usual $N=2$ critical string
in $D=2$ complex dimensions, {\it i.e.}\ four real dimensions.  The
existence of the complex structure which is required by the $N=2$ worldsheet
supersymmetry [4] makes the usual $N=2$ superstring rather unattractive
since it implies that the spacetime signature is $(2,2)$.  However, this
argument can be relaxed by introducing a background charge lying in a
certain direction without spoiling the local $N=2$ supersymmetry [3].
Indeed, the $N=2$ super-Virasoro algebra can be realised in terms of
arbitrary number of $N=2$ superfields and with background charges  fixed so
that the theory is critical. This gives rise to critical $N=2$ superstrings
in arbitrary numbers of complex dimensions.  The existence of background
charges can freeze one real coordinate, {\it i.e.} the momentum component in
that direction is constrained by the physical-state conditions to take some
specific values.  If one starts from a theory with one complex time and
chooses the background charge to lie in that direction, then the result of
the coordinate freezing is a theory that effectively has only one real time
direction.

       Beside the fact that the generalised $N=2$ superstrings make it
possible to describe the critical theories effectively in Minkowskian
spacetime signature, {\it i.e.}\ $(1,2p)$ with $p \ge 2$,  there are
fundamental differences between the generalised $N=2$ superstrings and the
usual one in $(2,2)$ spacetime.  The usual $N=2$ superstring is a
highly-degenerate theory.  The physical spectrum consists only of a massless
scalar which turns out to be the K\"ahler potential for the four dimensional
metric [1].  The corresponding interaction theory only involves upto
three-point functions [1].  Vanishing of four-point functions and beyond is
necessary for the consistency of the theory since higher-point functions
tend to produce mass poles which do not exist in the usual $N=2$
superstring.  Since the usual $N=2$ superstring turns out be a very
interesting theory which provides a consistent quantum theory of self-dual
gravity in four dimensions [1], it is most intriguing to study the more
general cases in order to uncover the full richness of the $N=2$
superstrings.

      It has been shown in [3] that the physical spectrum of the critical
$N=2$ superstrings in $D\ge 3$ complex dimensions comprises an infinite
tower of particles just like other string theories.  The no-ghost theorem
for the usual $N=2$ string have been proven in [5]. Norms of low-lying
physical states for generalised $N=2$ superstring are calculated in [3] and
the results indicate the unitarity of the theories.   In this paper we shall
first review the physical spectrum and then calculate the four-point
functions for the tachyonic state and show that they do not vanish in
general.  The mass poles from the gamma functions in the four-point
functions are consistent with the physical-state conditions.

\bigskip\bigskip
\noindent
{\bf 2. Physical Spectrum of Generalised $N=2$ Superstrings}
\bigskip\bigskip

        In this section, we shall review the discussion of ref.\ [3], where
the critical $N=2$ superstrings in $D\ge 3$ complex dimensions are
constructed. Since we only study $N=2$ superstrings in this paper, we shall
no longer repeat ``$N=2$'' unless there is ambiguity.  The discussion of the
physical spectrum applies to both open strings and close strings, for
simplicity, we shall only consider the holomorphic sector in this section.

      The super-Virasoro algebra is generated by a super energy-momentum
tensor which can be realised in terms of free chiral superfields
$$
\eqalign{
\Phi^+_{\mu}(z,\theta^+,\theta^-)&=\phi_{\mu}(z)+\sqrt2\theta^-
\psi_{\mu}(z)- \theta^+\theta^-\partial\phi_{\mu}(z)\ ,\cr
\Phi^-_{\mu}(z,\theta^+,\theta^-)&=\bar\phi_{\mu}(z)+\sqrt2\theta^+
\bar\psi_{\mu}(z)+\theta^+\theta^-\partial\bar\phi_{\mu}(z)\ ,\cr}
\eqno(2.1)
$$
where $\mu=(0,1,\ldots,D-1)$.  The superfields satisfy the chirality
conditions
$$
D^+\Phi^-_\mu=0=D^{-}\Phi^+_\mu\ ,\eqno(2.2)
$$
where
$$
D^{\pm}\equiv{\partial\over {\partial \theta^{\mp}}}+ \theta^{\pm}
\partial\ .\eqno(2.3)
$$
The operator-product expansions of these free superfields are given by
$$
\Phi^+_\mu(z_1, \theta^+_1, \theta^-_1)\Phi^-_\nu (z_2,\theta^+_2,
\theta^-_2)\sim -\eta_{\mu\nu}{\rm log}(Z_1-Z_2)\ ,\eqno(2.4)
$$
where
$$
Z_1-Z_2\equiv z_1-z_2+ \theta^+_1\theta^-_1+\theta^+_2\theta^-_2
-2\theta^+_1\theta^-_2\ .\eqno(2.5)
$$
The super energy-momentum tensor $T(z,\theta^+,\theta^-)$ can then be
written as
$$
T(z,\theta^+,\theta^-)=-\ft14 D^+\Phi^{+\mu}D^-\Phi^-_\mu + {\a\over 2}
\del\Phi^{+}_0-{\a\over 2}\partial\Phi^-_0\ .\eqno(2.6)
$$
Note that the background charge in expression (2.6) has been chosen to lie
in the $\mu=0$ direction for the reason explained in the introduction.
The central charge of this realisation is
$$
c=3(D-2\a^2)\ .\eqno(2.7)
$$
The anomaly-freedom condition $c=6$ therefore requires that
$$
\a^2=\ft12(D-2)\ .\eqno(2.8)
$$
We shall only consider in this paper the cases that $D\ge 3$ and thus it
follows from (2.8) that the background charge is real.

      The super energy-momentum tensor $T(z,\theta^+,\theta^-)$ may be
expanded in components as
$$
T(z,\theta^+,\theta^-)=\ft12 J(z)-\ft12\theta^+ G^-(z)+ \ft12\theta^-
G^+(z) + \theta^+\theta^- T(z)\ .\eqno(2.9)
$$
The component currents $(J, G^+, G^-, T)$ have conformal spins
$(1,\ft32,\ft32,2)$ measured by the energy-momentum tensor $T(z)$.  The
explicit forms of these component currents are given by
$$
\eqalign{
J&=-\psi^\mu\bar\psi_\mu - \a \del\phi_0 +\a \del\bar\phi_0 \ ,\cr
G^+&=\sqrt2(\del \bar\phi^\mu\psi_\mu- \a\del\psi_0)\ ,\cr
G^-&=\sqrt2(\del\phi^\mu\bar\psi_\mu - \a\del\bar\psi_0)\ , \cr
T&=\ft12\psi^\mu\del\bar\psi_\mu - \ft12\del\psi^\mu\bar\psi_\mu -
\del\phi^\mu\del\bar\phi_\mu + \ft12\a\del^2\phi_0 +
\ft12\a\del^2\bar\phi_0\ . \cr}\eqno(2.10)
$$

      A physical state $|p\big\rangle$ satisfies the physical-state
conditions
$$
\eqalign{
L_m\big|p\big\rangle&=0=J_m\big|p\big\rangle \qquad m\ge 0\ ,\cr
G^+_r\big|p\big\rangle&=0=G^-_r\big|p\big\rangle \qquad r>0\ .\cr}
\eqno(2.11)
$$
(Since the intercepts of $J_0$ and $L_0$ are zero for the $N=2$
super-conformal algebra, we include these in the physical-state conditions
(2.11).) Physical states can be constructed by acting on an
$SL(2,C)$-invariant vacuum $\big|0\big\rangle$ with physical operators
(ground-state operators) $P(z)$, {\it i.e.}\ $\big|p\big\rangle\equiv
P(0)\big|0\big\rangle$.  The physical operators take the form
$$
P(z)=R(z) e^{\beta\cdot\bar\phi+\bar\beta\cdot\phi}\ .\eqno(2.12)
$$
(Normal ordering is understood.)  The operators $R(z)$ can be classified by
their eigenvalues $q$ and $n$ under $J_0$ and $L_0$ respectively.  The
eigenvalue $q$ measures the fermion charge ($U(1)$ charge) of the operator
$R(z)$; each $\psi_\mu$ in a monomial in $R(z)$ contributes $+1$, each
${\bar\psi}_\mu$ contributes $-1$, and $\del\phi_\mu$ and $\del\bar\phi_\mu$
contribute 0. The eigenvalue $n$ measures the conformal dimension of the
operator $R(z)$, {\it i.e.}\ the level number. The corresponding vertex
operators, which are defined as total derivatives under the super
energy-moentum tensor, are given by
$$
V(z)=\ft12 (G^+_{\sss -\ft12}G^-_{\sss-\ft12}-G^-_{\sss -\ft12}
G^+_{\sss -\ft12})\,P(z)
\ ,\eqno(2.13)
$$

    At level $n=0$, $R$ is just the identity operator, with $q=0$, and
$P(z)$ is the ``tachyon" ground-state operator. At level $n=\ft12$, $R$
can be $\bar\xi_\mu\psi^\mu$, with $q=+1$; or $\xi_\mu\bar\psi^\mu$, with
$q=-1$. At level $n=1$, $q$ can be $-2,\ 0,\ +2$. In general, at level
$n$, $q$ takes the values
$$
q=-2n,\ -2n+2,\ \ldots,\ 2n-2,\ 2n\ .\eqno(2.14)
$$
For physical states with level number $n$ and fermion
charge $q$, the $J_0$ and $L_0$ constraints in (2.11) give
$$
\eqalignno{
J_0:\qquad  0&=q+\a (\beta_0-\bar\beta_0)\,&(2.15a)\cr
L_0:\qquad 0&=n -\beta^\mu \bar\beta_\mu +\ft12 \a (\beta_0+\bar\beta_0)
\ .&(2.15b)\cr}
$$
In the usual discussion of the $N=2$ superstring, for which $D=2$ and hence
from (2.8) the background charge $\a$ is zero, equation (2.15$a$) implies
that for all physical states the fermionic charge of $R(z)$ must be zero.
This implies that in this case there are no physical states occurring at
levels with $n$ a half-integer.  (In fact, as discussed in [1,5,6], all the
higher-level  states satisfying conditions (2.15$a,b$) are longitudinal, and
hence have no physical degrees of freedom.)  It follows from equation
(2.15$a$) that the real momentum component $\bar\beta_0-\beta_0$ is
``frozen'' to the value
$$
\bar\beta_0-\beta_0= {q\over \a}\ .\eqno(2.16)
$$
Thus the real time direction $(\phi_0-\bar\phi_0)$ is not a
physical-observable coordinate; effectively one is left with one real time
coordinate $(\phi_0+\bar\phi_0)$.  (Momentum-freezing is a genuine
phenomenon for $W$-string theories with non-linear local symmetry algebras,
namely $W$ algebras [7]. It is remarkable that such phenomenon occurs also
in the linear superstrings.)

\bigskip\bigskip
\noindent
{\bf 3. Four-point Function of Tachyonic physical operators}
\bigskip\bigskip

      Having obtained a complete set of physical operators in section 2, we
shall turn our attention to interaction theories.  We shall concentrate on
four-point function of the tachyonic physical state, {\it i.e.} level-zero
state.  The corresponding vertex operator can be evaluated from (2.13) as
$$
V_0(z)=\big(\beta\cdot\del\bar\phi-\bar\beta\cdot\del\phi-2(\beta\cdot\bar
\psi)(\bar\beta\cdot\psi)\big)e^{\beta\cdot\bar\phi+\bar\beta\cdot\phi}\ ,
\eqno(3.1)
$$
where ``$\cdot$'' stands for the contraction of $\mu$ indices.  In section 2
we choose the component form to construct the physical spectrum.  It is more
convenient to work with the superfield language to calculate the $n$-point
functions.  In superfield form, the vertex operators in (2.13) can be
expressed equivalently as
$$
\eqalign{
V(z)&=\int d^2\theta\, V(z,\theta^{\pm}) \cr
    &=\int d^2\theta\,  R(D^+\Phi^+,D^-\Phi^-)e^{\bar\beta\cdot\Phi^+ +
\beta\cdot\Phi^-}\ ,\cr}\eqno(3.2)
$$
where $R(D^+\Phi^+,D^-\Phi^-)$ is the corresponding super differential
polynomial on superfields $\Phi^+_\mu$ and $\Phi^-_\mu$.  When $R$ is just
the identity operator, it gives rise to the vertex operator for the
tachyonic state in superfield form
$$
V_0(z,\theta^{\pm})=e^{\beta\cdot\Phi^+ +
\bar\beta\cdot\Phi^-}\ .\eqno(3.3)
$$
Integrating out the $\theta^{\pm}$ coordinates leads to the vertex operator
(3.1).  The free superfields $\Phi^+_\mu$ and $\Phi^-_\mu$ satisfy the OPEs
given by (2.4) and (2.5).   The n-point correlation functions are then given
by
$$
\big\langle V_0(z_1,\theta^{\pm}_1)V_0(z_2,\theta^{\pm}_2)\ldots
V_0(z_n,\theta^{\pm}_n)\big\rangle=\prod_{i<j}(Z_i-Z_j)^{-(\beta_i
\cdot\bar\beta_j+\bar\beta_i\cdot\beta_j)}\ ,\eqno(3.4)
$$
Where $(Z_i-Z_j)$ is defined  in (2.5).  The $\beta^\mu_i$ and
$\bar\beta^\mu_i$ in (3.4)  satisfy the momentum-conservation law
$$
\sum^n_i\beta^\mu_i=-\alpha^\mu \qquad \sum^n_i\bar\beta^\mu_i=-
\alpha^\mu\ ,\eqno(3.5)
$$
where $\alpha_\mu\equiv(\a, 0,0,\ldots, 0)$.  The momentum-conservation law
is modified in $\mu=0$ direction by the background charge.

       So far our discussion applies to both open strings and closed
strings. We shall calculate the four-point amplitude only for closed
superstrings.  Including the anti-holomorphic sector, the tachyonic vertex
operator for the closed string is given by
$$
V_0(z,\bar z,\theta^{\pm},\bar\theta^{\pm})=e^{\bar\beta\cdot\big
(\Phi^+(z,\theta^{\pm})+\bar\Phi(\bar z,\bar\theta^{\pm})\big) +
\beta\cdot\big(\Phi^-(z,\theta^{\pm})+\bar\Phi^-(\bar z,
\bar\theta^{\pm})\big)}\ ,\eqno(3.6)
$$
where $\bar\Phi^{\pm}_\mu(\bar z,\bar\theta^{\pm})$ are the anti-holomorphic
superfields and $(\bar z, \bar \theta^{\pm})$ are anti-holomorphic
super-coordinates.   To calculate the $n$-point amplitudes, one need to
integrate out all the super-coordinates on the super Riemann sphere.  Since
there are three conformal Killing vectors and two super-conformal Killing
spinors, we can fix three bosonic coordinates to $\infty$, $1$, and $0$, and
set two pairs of fermionic coordinates to zero.  Thus the three-point
function is given by
$$
\eqalign{
A_3&=\int d^2\theta d^2\bar\theta\,\big\langle V_0(z_1=\infty,0,0)
V_0(z_2=1,\theta^{\pm},\bar\theta^{\pm})V_0(z_3=0,0,0)\big\rangle\cr
&=(\beta_2\cdot\bar\beta_3-\bar\beta_2\cdot\beta_3)^2\ .\cr
}\eqno(3.7)
$$

      Similarly, the four-point function is given by
$$
\eqalign{
A_4&=\int d^2zd^2\theta_2 d^2\bar\theta_2d^2\theta_3 d^2\bar\theta_3\,
\big\langle V_0(\infty,0,0)V_0(1, \theta^{\pm}_2,\bar\theta^{\pm}_2)
V_0(z,\theta^{\pm}_3, \bar\theta^{\pm}_3)V_0(0,0,0)\big\rangle\cr
&=\int d^2z\,\big |{t(t+1)\over (1-z)^2}+{c_{12}c_{34}\over z} +
{c_{23}c_{41}\over (1-z)}\big |^2 \big |z\big |^{-2s}\big |1-z\big |^{-2t}
\ ,\cr}\eqno(3.8)
$$
where $c_{ij}=\bar\beta_i\beta_j-\beta_i\bar\beta_j$ is a square-root of the
three-point function given in (3.7) and $s=\bar\beta_1\cdot\beta_2+\beta_1
\cdot\bar\beta_2$, $t=\bar\beta_2\cdot\beta_3+\beta_2\cdot\bar\beta_3$  and
$u=\bar\beta_1\cdot\beta_3+\beta_1\cdot\bar\beta_3$.  Using our definition
of Mandelstam variables   it is easy to check that even with the presence of
the background charge, the evaluation of the four-point function (3.8) gives
the same form  as the one of the usual D=2 critical superstring in [1].  The
result is
$$
A_4=\pi F^2{\Gamma(1-s)\Gamma(1-t)\Gamma(1-u)\over {\Gamma(s)\Gamma(t)
\Gamma(u)}}\ ,\eqno(3.9)
$$
and the identity
$$
s+t+u=0\eqno(3.10)
$$
still holds in the presence of the background charge.   The prefactor $F$ in
(3.9) takes the same form as the one in the usual superstring
$$
F=1-{c_{12}c_{34}\over{su}}-{c_{23}c_{41}\over{tu}}\ .\eqno(3.11)
$$
Using the modified momentum-conservation law (3.5) and the intercept
condition (2.15$a$,$b$), one can rewrite $c_{34}=c_{13}+c_{23}$ and
$c_{41}=c_{12}+c_{13}$. Substituting these into (3.11) and using (3.10), one
has
$$
F={4\over {stu}}\Big((\beta_1\cdot\bar\beta_2)(\beta_2\cdot\bar\beta_3)
(\beta_3\cdot\bar\beta_1)+(\bar\beta_1\cdot\beta_2)(\bar\beta_2\cdot\beta_3)
(\bar\beta_3\cdot\beta_1)\Big)\ .\eqno(3.12)
$$

      The form of the four-point function is identical for the usual
super-string and the generalised superstrings.  However the mass-shell
conditions and the spacetime dimensions are different.  These differences
play important r\^oles of the final result of the four-point functions.  In
the case of usual superstring where spacetime signature is $(2,2)$ and hence
there is no background charge, the prefactor $F$ vanishes due to the
on-shell condition of the external momenta
$\beta_{i}\cdot\bar{\beta}_{i}=0$.   For the generalised superstring, the
prefactor is not zero in general.  For example, when
$\beta_1=\beta_2=\beta_3=\beta$ and $\bar\beta_1=\bar\beta_2=\bar\beta_3=
\bar\beta$, the prefactor $F$ is proportional to $(\beta\cdot\bar\beta)^3$
which does not vanish since the mass-shell condition is
$\beta\cdot\bar\beta=\a\beta_0$.

      The gamma functions in (3.9) will produce mass-poles in the scattering
amplitudes. For $s$-channel the singularity occurs when $s=n\ge 1$, which
gives exactly the intercept conditions (2.15$a$,$b$) with $q=0$ and $n$
integers.  Similar analysis applies to $t$- and $u$-channels as well. Since
the fermion charge of the tachyonic vertex operator is zero, owing to the
charge conservation, the exchange-particle channels only involve the
physical states with vanishing fermion charge $q=0$.   In fact, from (2.16),
the the fermion charge is proportional to the momentum.  The charge
conservation originates from the momentum conservation.

\bigskip\bigskip
\noindent
{\bf 4. Conclusions}
\bigskip\bigskip

       In this paper, we have looked at details of the generalised critical
$N=2$ superstrings in $(1,2p)$ spacetime signature.  We calculate the
four-point function of the theories. The form of the four-point function is
similar to the one of the usual $N=2$ superstring.  However, since the
spacetime signatures are different and the external momenta subject to
different on-shell conditions, the final implications of the four-point
function calculation are different.  The four-point function vanishes in the
usual $N=2$ superstring, while it is non-zero  in the generalised $N=2$
superstrings.  The mass poles in the gamma functions only involve the
physical states with vanishing fermion charge, which is consistent with the
charge conservation the theories possess.

     The charge conservation however does not exclude the possibility of
propagation of $q$-charged states in loops.  It can be easily checked that
the  momentum conservation constraint (3.5) applied to the scattering
process of  three physical states with charges $(q,-q,0)$ does not exclude
the possibility of propagating the $q\neq 0$ states in loops. However  we
can  compactify the frozen time coordinate coupled to
$(\bar{\beta}_{0}-\beta_{0})$ momentum on a circle with radius  $R\neq
{n\over m} \alpha_{0}$, where $n,m$ are natural numbers.  The quantised
momentum satisfies $\alpha_{0}(\bar{\beta}_{0}-\beta_{0})\neq m$ ($m$
integers except for zero). The constraint that  $q$ charge has to be integer
for a physical state then automaticaly reduces the physical states space to
the states with $q=0$. This condition also excludes all possible tachyons
from the physical states spectrum of the generalised N=2 superstrings [3].

\bigskip\bigskip
\bigskip\bigskip
\centerline{\bf ACKNOWLEDGEMENT}
\bigskip

      We are obliged to the TASI-92 conference at Boulder, CO. for the
hospitality during the course of this work. J.B. would like to
thank Emil Martinec for helpful discussions and comments.

\bigskip\bigskip
\centerline{\bf REFERENCES}
\frenchspacing
\bigskip

\item{[1]}H. Ooguri and C. Vafa, {\sl Nucl. Phys.} {\bf 361B} (1991) 469.

\item{[2]}H. Ooguri and C. Vafa, {\sl Mod. Phys. Lett.} {\bf A5}
 (1990) 1389.

\item{[3]}H.\ Lu, C.N.\ Pope, X.J.\ Wang, K.W.\ Xu, ``$N=2$ Superstrings
with $(1,2m)$ Spacetime Signature,''  preprint CTP TAMU-99/91, to appear
in {\sl Phys. lett.} {\bf B}.

\item{[4]}L. Alvarez-Gaum\'e and D.Z. Freedman, {\sl Comm. Math. Phys.}
 {\bf 80} (1981) 443.

\item{[5]}J. Bie\'nkowska, {\sl Phys. Lett.} {\bf 281B} (1992) 59.

\item{[6]}M.B.\ Green, J.H.\ Schwarz and E.\ Witten, ``Superstring Theory,''
(CUP 1987).

\item{[7]}C.N.\ Pope, L.J.\ Romans and K.S.\ Stelle, {\sl Phys. Lett.}
{\bf 268B} (1991) 167; {\sl Phys. lett.} {\bf 269B}(1991) 287;\nl
H.\ Lu, C.N.\ Pope, S.\ Schrans and K.W.\ Xu,``The Complete Spectrum of the
$W_N$ String,'' preprint CTP TAMU-5/91, KUL-TF-92/1, to appear in {\sl
Nucl. Phys.} {\bf B}.

\end